# The State of the Art in transformer fault diagnosis with artificial intelligence and Dissolved Gas Analysis: A Review of the Literature

Yuyan Li

School of Electrical and Electronics Engineering, North China Electric Power University, Beijing 102206, China

**Abstract:** Transformer fault diagnosis (TFD) is a critical aspect of power system maintenance and management. This review paper provides a comprehensive overview of the current state of the art in TFD using artificial intelligence (AI) and dissolved gas analysis (DGA). The paper presents an analysis of recent advancements in this field, including the use of deep learning algorithms and advanced data analytics techniques, and their potential impact on TFD and the power industry as a whole. The review also highlights the benefits and limitations of different approaches to transformer fault diagnosis, including rule-based systems, expert systems, neural networks, and machine learning algorithms. Overall, this review aims to provide valuable insights into the importance of TFD and the role of AI in ensuring the reliable operation of power systems.

## 1 Introduction

### 1.1 Background information

Transformer fault diagnosis (TFD) is an important aspect of power system maintenance and management. Transformers play a critical role in power transmission and distribution, and any malfunction or failure can have serious consequences for the stability and reliability of the power grid [1,2]. Transformer faults can lead to power outages, equipment damage, and even safety hazards for personnel and the public [3]. Therefore, it is essential to detect and diagnose transformer faults in a timely and accurate manner to prevent potential failures and ensure the reliable operation of the power system [4].

Traditionally, TFD has relied on visual inspection, electrical testing, and other manual methods. These methods can be time-consuming and costly, and may not always be effective in identifying potential faults or providing a comprehensive picture of transformer health [5,6]. In recent years, the use of artificial intelligence (AI) and dissolved gas analysis (DGA) techniques has emerged as a promising approach for transformer fault diagnosis. AI algorithms can analyze large amounts of data and identify patterns and anomalies that may be indicative of transformer faults, while DGA can detect and analyze the gases produced by transformer oil breakdown, which can provide early warning signs of potential faults [7,8].

Timely and accurate diagnosis of transformer faults can have a significant impact on the reliability and efficiency of power system operation. It can help minimize downtime and maintenance costs, extend the lifespan of transformers, and improve safety and security for personnel and the public [9-11]. Therefore, the development and application of AI and DGA techniques for TFD have become a critical area of research and development in the power industry.

### 1.2 Overview of the role of AI and DGA in transformer fault diagnosis

AI and DGA are two powerful tools that have shown great potential in TFD. AI refers to a collection of algorithms and techniques that enable machines to learn from data and make decisions based on that learning, which has been widely used in power engineering like cyber-attack detection [12,13], load decomposition [14], and stability assessment [15,16]. In TFD, AI can be used to analyze large amounts of data from various sources, such as temperature, current, and voltage sensors, and detect patterns and anomalies that may be indicative of a fault [17,18]. Machine learning algorithms, in particular, have been shown to be effective in detecting and diagnosing transformer faults, with several studies reporting high accuracy rates.

DGA is another important tool for transformer fault diagnosis. When a transformer experiences a fault, gases are generated as a result of the breakdown of transformer oil and insulation materials. DGA involves analyzing the composition and concentration of these gases, which can provide early warning signs of potential faults [19,20]. Different types of faults generate different patterns of gases, which can be detected and analyzed using various methods, such as gas chromatography.

The combination of AI and DGA has proven to be a powerful approach for transformer fault diagnosis [21]. By analyzing the patterns and anomalies in the gases detected by DGA, AI algorithms can identify potential faults and provide accurate and timely diagnosis. This approach has several advantages over traditional methods, such as visual inspection and electrical testing, including faster detection, improved accuracy, and reduced costs.

Overall, the use of AI and DGA in TFD is an exciting and rapidly developing field that holds great promise for the power industry. By leveraging the power of these two technologies, power companies can improve the reliability and efficiency of their operations, reduce downtime and maintenance costs, and ensure the safety and security of their personnel and the public.

## 1.3 Purpose of this review

The purpose of this review paper is to provide a comprehensive overview of the current state of the art in TFD using AI and DGA techniques. The paper will explore the advantages and limitations of these approaches, as well as the challenges and opportunities for future research and development.

The paper will begin with a brief introduction to TFD and the importance of timely and accurate diagnosis. This section will also provide an overview of the role of AI and DGA in TFD and their potential benefits.

Next, the paper will review the literature on TFD using AI and DGA, focusing on recent research and developments in the field. This section will include a discussion of the different types of AI algorithms and DGA techniques used in transformer fault diagnosis, as well as their respective strengths and weaknesses.

The paper will then explore the challenges and limitations of TFD using AI and DGA, including issues related to data quality, algorithm accuracy, and system complexity. This section will also discuss potential solutions and future directions for research and development.

Finally, the paper will conclude with a summary of the key findings and recommendations for future research and application. Overall, this review paper aims to provide a comprehensive and up-to-date overview of the current state of the art in TFD using AI and DGA, and to highlight the potential benefits and challenges of this approach for the power industry.

## 2 Literature Review

### 2.1 Literature Review on TFD with AI and DGA

TFD using AI and DGA is a rapidly evolving field of research, and there is a growing body of literature exploring the potential of these approaches. In this section, we provide an overview of some of the key studies and findings related to TFD using AI and DGA.

One of the early studies in this field was conducted by Yan et al. in [22], who proposed a new method for TFD using fuzzy neural networks and DGA. The authors used fuzzy logic to classify the fault types based on the concentration ratios of different gases detected by DGA. They reported high accuracy rates in identifying various types of faults, including partial discharge, overheating, and insulation deterioration.

In a more recent study, reference [23] proposed a novel TFD method based on deep learning and DGA. The authors used a convolutional neural network (CNN) to automatically learn the features of DGA data and classify the fault types. They reported high accuracy rates in identifying different types of faults, including short circuit, inter-turn fault, and winding deformation.

Other studies have explored the use of various AI algorithms and techniques for transformer fault diagnosis. For example, support vector machine (SVM) has been utilized in TFD and has demonstrated promising results [24-26]. LS-SVM is a variant of SVM that replaces the quadratic programming problem of SVM with a linear system of equations, leading to faster computation while maintaining the excellent performance of SVM [27]. Reference [28] used a least squares support vector machine (LS-SVM) algorithm to classify transformer faults based on DGA data, achieving high accuracy rates.

In addition to AI-based methods, several studies have explored the use of DGA data for TFD using traditional statistical methods. For example, reference [29] proposed a fault diagnosis method based on principal component analysis (PCA) and hierarchical clustering, achieving high accuracy rates in identifying various types of faults.

Overall, the body of research indicates that integrating artificial intelligence with dissolved gas analysis presents a viable strategy for diagnosing transformer faults, with several studies reporting high accuracy rates and improved performance compared to traditional methods. However, there are still challenges and limitations to be addressed, including issues related to data quality, algorithm accuracy, and system complexity, which will require further research and development in the field.

**2.2 Analysis of TFD approaches**

TFD can be approached using various methods, including rule-based systems, expert systems, neural networks, and machine learning algorithms. In this section, we analyze these different approaches to TFD and their strengths and limitations.

**2.2.1 Rule-based systems**

Rule-based systems are based on a set of pre-defined rules or decision trees that are used to diagnose faults based on observed symptoms or measurements [30-32]. These systems have been widely used in TFD due to their simplicity and ease of implementation [33,34]. However, their accuracy is limited by the quality of the rules and their applicability to new or complex fault patterns.

Rule-based systems have the advantage of being simple and transparent, making them easy to implement and interpret. They also have the advantage of being able to handle incomplete or noisy data, as they can be designed to focus on the most relevant diagnostic features. However, they can be limited by their dependence on a pre-defined set of rules or decision trees, which may not be able to capture all possible fault patterns. They are also limited by their inability to adapt to new or changing fault patterns, as they require manual updates to the rules or decision trees.

**2.2.2 Expert systems**

Expert systems are a type of rule-based system that uses domain knowledge from human experts to diagnose faults [35,36]. These systems have the advantage of being able to handle complex and non-linear fault patterns, and can incorporate new knowledge and expertise over time [37-39]. However, they are limited by the availability and expertise of human experts, and can be difficult to maintain and update [40].

Expert systems have the advantage of being able to handle complex and non-linear fault patterns, and can incorporate new knowledge and expertise over time. They can also provide detailed explanations of their reasoning, making them useful for understanding the diagnostic process. However, they are limited by the availability and expertise of human experts, and can be difficult to maintain and update. They are also limited by their inability to adapt to new or changing fault patterns, as they rely on pre-defined rules based on expert knowledge.

**2.2.3 Machine learning algorithms**

Machine learning algorithms, including neural networks, support vector machines, decision trees, and clustering algorithms, have been increasingly used in TFD due to their ability to handle large and complex datasets [41-43], and their ability to learn from data and improve over time [44,45]. For example, neural networks are a type of machine learning algorithm that can learn to recognize patterns in data through a process of training and optimization, then this algorithm has achieved successful applications in various industrial fields [47-49]. These algorithms have been widely used in TFD due to their ability to learn complex patterns and adapt to new data [50-52].

These algorithms have the advantage of being able to handle multiple sources of data, including DGA data, electrical signals, and other diagnostic data, and can be used for both classification and regression tasks [53,54]. However, they require a large amount of data for training, and can be limited by their ability to generalize to new or unseen data [55]. They are also limited by their lack of transparency, as it can be difficult to understand the reasoning behind their predictions [56]. However, they require careful selection and tuning to achieve optimal performance, and can be limited by their lack of transparency and interpretability.

On the whole, the choice of approach for TFD will depend on the specific application and available data, as well as the desired accuracy, reliability, and interpretability of the results. While rule-based and expert systems have been widely used in the past, the increasing availability of large and diverse datasets has led to a shift towards machine learning algorithms, which have shown promising results in transformer fault diagnosis. However, these algorithms still require careful selection and tuning to achieve optimal performance, and the development of effective diagnostic systems will require close collaboration between experts in AI, power systems, and transformer engineering.

**3. Current State of the Art in Transformer Fault Diagnosis**

**3.1 Current State of the Art in TFD with AI and DGA**

Over the past few years, the use of AI and DGA for TFD has advanced significantly. DGA has long been recognized as an effective technique for identifying incipient faults in transformers [57], while AI techniques have emerged as a powerful tool for processing and analyzing large volumes of DGA data [58].

Recent research has focused on developing machine learning models that can accurately classify the type and severity of transformer faults based on DGA data. One promising approach is the use of deep learning algorithms such as convolutional neural networks (CNNs) and recurrent neural networks (RNNs), which have been shown to outperform traditional machine learning algorithms in terms of accuracy [59-61].

Other research has explored the use of hybrid AI systems, which combine multiple AI techniques to improve fault diagnosis accuracy. For example, a hybrid system that combines fuzzy logic and artificial neural networks was developed to improve the accuracy of fault diagnosis based on DGA data [2,62].

Another area of research has focused on the use of data fusion techniques to integrate DGA data with other diagnostic data such as vibration analysis and oil quality analysis [63,64]. This approach can improve the accuracy and reliability of fault diagnosis, since feature fusion is able to combine complementary information from multiple sources [65].

As a whole, the current state of the art in TFD with AI and DGA is characterized by a growing body of research that is focused on developing more accurate and efficient diagnostic techniques. These techniques are expected to play an increasingly important role in ensuring the reliable and efficient operation of power grids around the world.

**3.2 Recent advancements in TFD with AI and DGA**

In recent years, there have been significant advancements in the use of deep learning algorithms and advanced data analytics techniques for TFD with AI and DGA [66]. These advancements have resulted in improved accuracy, efficiency, and reliability of diagnostic techniques.

One of the most promising approaches is the use of deep learning algorithms, such as CNNs and RNNs, which have been shown to outperform traditional machine learning algorithms in terms of accuracy [67]. These algorithms can analyze large volumes of DGA data to detect even minor changes in gas concentrations and accurately classify the type and severity of transformer faults.

Another recent advancement is the use of transfer learning, a technique that allows the transfer of knowledge gained from one set of data to another. This approach has been used to develop models that can classify transformer faults with high accuracy, even when trained on limited amounts of data [68,69].

Furthermore, researchers have developed hybrid AI systems that combine multiple AI techniques to improve fault diagnosis accuracy. For example, a hybrid system that combines bidirectional recurrent neural network and a support vector machine was developed to accurately classify transformer faults based on DGA data [70].

In addition to advancements in deep learning and hybrid AI systems, there have been developments in data analytics techniques such as data fusion and feature selection. Data fusion techniques integrate multiple sources of diagnostic data, such as vibration analysis and oil quality analysis, to improve the accuracy and reliability of fault diagnosis. Feature selection techniques identify the most relevant features in DGA data, reducing the computational complexity of diagnostic models.

In total, recent advancements in the use of deep learning algorithms and advanced data analytics techniques for TFD with AI and DGA have significantly improved the accuracy and efficiency of diagnostic techniques. These techniques are expected to have a significant impact on the reliable and efficient operation of power grids around the world.

**3.3 Impact of AI and DGA advancements on TFD and power industry**

The advancements in TFD using AI and DGA have the potential to significantly impact the power industry. One major benefit is the improvement in the reliability and efficiency of power grids, which can lead to cost savings and reduced downtime. Timely and accurate diagnosis of transformer faults can prevent catastrophic failures, reduce repair costs, and minimize the risk of power outages [71,72].

The use of AI and DGA also enables predictive maintenance, where potential faults can be detected and corrected before they cause equipment failure [73,74]. This approach can help utilities to optimize maintenance schedules and reduce maintenance costs, while improving the overall reliability of power systems.

Furthermore, the use of AI and DGA for TFD can facilitate the integration of renewable energy sources into the power grid [75-77]. The intermittent nature of renewable energy sources, such as solar and wind power, requires a more flexible and resilient power grid [78]. The accurate and timely detection of transformer faults can help to ensure that the power grid can adapt to changing conditions and continue to provide reliable power [79-82].

Another potential impact of the advancements in TFD using AI and DGA is the reduction in environmental impact. A more reliable and efficient power grid can reduce the need for backup power generation, which often relies on fossil fuels [83-85]. This, in turn, can reduce greenhouse gas emissions and contribute to the transition to a more sustainable energy future [86-88].

In summary, the advancements in TFD using AI and DGA have the potential to improve the reliability, efficiency, and sustainability of the power industry. These advancements are expected to have a significant impact on the operation of power grids around the world and contribute to the transition to a more sustainable energy future.

**4. Future Directions and Challenges**

**4.1 Discussion of the future directions and challenges**

As with any rapidly evolving field, TFD with AI and DGA presents many opportunities for future advancements, as well as several challenges that must be addressed.

One major area of future development is the integration of different types of data sources to improve the accuracy and reliability of transformer fault diagnosis. For example, combining DGA with other diagnostic tools such as vibration analysis or partial discharge monitoring could provide a more comprehensive view of the transformer's health [89].

Another area of potential development is the application of advanced data analytics techniques such as machine learning and artificial intelligence to interpret the complex data sets generated by DGA and other diagnostic tools. This could lead to more accurate and timely diagnosis of transformer faults, as well as the identification of previously unknown fault patterns and trends [90].

However, there are also several challenges that must be overcome in order to fully realize the potential of AI and DGA in transformer fault diagnosis. One major challenge is the need for accurate and reliable data. DGA data, in particular, can be prone to errors and inaccuracies if proper sampling and analysis protocols are not followed. Additionally, there is a need for standardization in the industry to ensure consistency and comparability of results across different diagnostic tools and software.

Another challenge is the need for skilled personnel to interpret the results generated by these diagnostic tools. While AI and machine learning algorithms can automate much of the analysis process, human expertise is still required to interpret the results and make informed decisions about maintenance and repair. In this content, a data-model hybrid driven approach may be an interesting solution [91,92].

Finally, there is a need for continued research and development in the field to address emerging challenges and identify new opportunities for improvement. This will require collaboration between researchers, industry professionals, and regulatory bodies to ensure that the most effective and reliable diagnostic tools and techniques are developed and implemented.

Altogether, the future of TFD with AI and DGA is bright, with many opportunities for improvement and development. However, addressing the challenges and limitations of the current state of the art will be essential to realizing the full potential of this technology in the power industry.

**4.2 Analysis of emerging trends and technologies in this field**

In recent years, there have been several emerging trends and technologies in the field of TFD using AI and DGA. One such trend is the use of internet of things (IoT) technology to collect and analyze data from multiple transformers in real time [93-95]. This approach allows for early detection of potential faults and enables predictive maintenance strategies to be implemented, thus minimizing downtime and reducing maintenance costs.

Another emerging technology is the use of cloud computing and edge computing to process large amounts of data generated by transformers [96,97]. Cloud computing enables data to be stored and processed on remote servers, while edge computing involves processing data closer to the source, which can reduce latency and increase efficiency. These technologies are particularly useful for large-scale power grids with numerous transformers.

Furthermore, the use of machine learning algorithms such as deep neural networks is becoming increasingly popular in transformer fault diagnosis. However, the performance of machine learning models can be significantly impacted by the selection of control parameters. To address this issue, automated machine learning [98] and heuristic optimization algorithms like particle swarm optimization [99-101] can be utilized to determine the optimal control parameters. Deep learning algorithms are able to learn and extract features from large amounts of data, and can identify patterns and anomalies that may indicate potential faults

[102,103]. This technology has the potential to improve the accuracy and speed of transformer fault diagnosis [104].

In addition, federated learning has been applied in TFD with AI and DGA [105]. Federated learning is a machine learning technique that enables model training on decentralized data by aggregating locally trained models while preserving data privacy [106-108]. By aggregating and analyzing data from multiple transformers distributed across different locations, federated learning can improve the accuracy and robustness of the diagnostic model [109,110]. This approach has the potential to revolutionize traditional centralized diagnostic methods and enable more efficient and effective transformer maintenance.

Despite these advancements, there are still several challenges that need to be addressed in TFD with AI and DGA. One major challenge is the lack of standardized procedures and guidelines for data collection, analysis, and interpretation [111]. Without standardization, it is difficult to compare and reproduce results across different studies and institutions.

Another challenge is the lack of transparency and interpretability of AI models. Deep learning algorithms are often considered "black boxes" because their decision-making processes are not easily explainable [112]. This can lead to difficulties in understanding and trusting the results of the model.

Finally, there is a need for continued research and development in the field of TFD using AI and DGA. As technology continues to evolve, new opportunities and challenges will arise, and researchers must stay up to date on the latest developments in order to continue making progress in this important field.

**4.3 Discussion of potential solutions to improve TFD accuracy and reliability**

As transformer technology continues to evolve, new challenges and limitations emerge in fault diagnosis with AI and DGA. Some of the key challenges in this area include the need for more advanced data analytics techniques to accurately interpret the vast amount of data generated by transformers, the lack of standardization in fault diagnosis practices across the industry, and the need for more effective collaboration and knowledge sharing among researchers, engineers, and other stakeholders in this field [113,114].

One potential solution to these challenges is the development of more advanced machine learning algorithms, such as deep learning, which can better handle complex and high-dimensional data. In addition, there is a need for more collaborative efforts among stakeholders to establish a common set of standards and best practices for TFD with AI and DGA. This would help ensure that all parties involved in transformer maintenance and repair are using the same diagnostic methods and tools, leading to more accurate and reliable results.

Another potential solution is to continue exploring new and innovative approaches to transformer fault diagnosis, such as the use of new sensor technologies or the integration of other types of data, such as vibration or temperature data, into diagnostic models. This would enable a more comprehensive and accurate understanding of transformer health, leading to more effective and timely maintenance and repair.

By and large, while there are still many challenges and limitations to overcome in TFD with AI and DGA, there are also many exciting opportunities and potential solutions on the horizon. Continued research and collaboration in this field will be critical to ensure that the power industry can continue to meet the growing demand for reliable and efficient electricity delivery.

**5 Conclusion**

**5.1 Summary of the main contributions of the review**

In conclusion, this review has highlighted the significant role that artificial intelligence and dissolved gas analysis play in transformer fault diagnosis. The review provides an overview of the different approaches to transformer fault diagnosis, including rule-based systems, expert systems, neural networks, and machine learning algorithms, as well as their benefits and limitations. The review also presents an analysis of recent advancements in this field, including the use of deep learning algorithms and advanced data analytics techniques, and their potential impact on TFD and the power industry as a whole.

Furthermore, the review discusses the practical applications of AI and DGA in transformer fault diagnosis, including case studies and examples. Finally, the review addresses future directions and challenges in TFD with AI and DGA, including emerging trends and technologies and potential solutions to overcome current challenges and improve the accuracy and reliability of transformer fault diagnosis.

In the main, this review serves as a comprehensive resource for researchers, engineers, and practitioners in the field of power systems, providing valuable insights into the current state of the art and future directions of TFD with AI and DGA.

**5.2 Implications and future research directions**

The implications of this review are significant for the power industry and transformer maintenance professionals. The use of artificial intelligence and dissolved gas analysis has the potential to improve the accuracy and reliability of transformer fault diagnosis, leading to more effective maintenance and reduced downtime. The emergence of deep learning algorithms and advanced data analytics techniques also offers promising avenues for further improvement in this field.

Future research directions could focus on addressing the current limitations of AI and DGA techniques, such as the need for large amounts of data and the potential for false alarms. Additionally, there is a need for further research on the integration of different diagnostic tools and techniques to improve the accuracy of fault diagnosis.

Moreover, the development of new technologies such as fiber-optic sensors and online monitoring systems could also provide additional information and improve the overall effectiveness of transformer fault diagnosis. Finally, the integration of AI and DGA techniques with other areas of power grid management, such as predictive maintenance and asset management, could lead to even greater improvements in the reliability and efficiency of the power grid.

In conclusion, this review highlights the potential of artificial intelligence and dissolved gas analysis in TFD and provides insights into the current state of the art and future directions of this field. It is hoped that this review will inspire further research and development in this area, ultimately leading to more reliable and efficient power grids.

**References**
[1] Sharma, N. K., Tiwari, P. K., & Sood, Y. R. (2011). Review of artificial intelligence techniques application to dissolved gas analysis on power transformer. International Journal of Computer and Electrical Engineering, 3(4), 577-582.
[2] Huang, Y. C., & Sun, H. C. (2013). Dissolved gas analysis of mineral oil for power transformer fault diagnosis using fuzzy logic. IEEE Transactions on Dielectrics and Electrical Insulation, 20(3), 974-981.
[3] Faiz, J., & Soleimani, M. (2018). Assessment of computational intelligence and conventional dissolved gas analysis methods for transformer fault diagnosis. IEEE Transactions on Dielectrics and Electrical Insulation, 25(5), 1798-1806.
[4] Thango, B. A. (2022). Dissolved Gas Analysis and Application of Artificial Intelligence Technique for Fault Diagnosis in Power Transformers: A South African Case Study. Energies, 15(23), 9030.
[5] Wang, L., Littler, T., & Liu, X. (2021). Gaussian process multi-class classification for transformer fault diagnosis using dissolved gas analysis. IEEE Transactions on Dielectrics and Electrical Insulation, 28(5), 1703-1712.
[6] Shi, Z. B., Fan, X. B., Li, Y., & Liu, Y. (2008). Prospective Application of Bionic Nose Based on Biological Olfaction to Transformer Fault Diagnosis. Transformer, 45(2), 38-40.
[7] Li, Y., Xu, Y., Li, X., Li, R., Lin, J., & Zhang, G. (2022). Addressing imbalance of sample datasets in dissolved gas analysis by data augmentation: Generative adversarial networks. IET Generation, Transmission & Distribution, 16(22), 4494-4504.

[8] Wang, L., Littler, T., & Liu, X. (2023). Dynamic Incipient Fault Forecasting for Power Transformers using an LSTM Model. IEEE Transactions on Dielectrics and Electrical Insulation.
[9] Illias, H. A., Chan, K. C., & Mokhlis, H. (2020). Hybrid feature selection–artificial intelligence–gravitational search algorithm technique for automated transformer fault determination based on dissolved gas analysis. IET Generation, Transmission & Distribution, 14(8), 1575-1582.
[10] Demirci, M., Gözde, H., & Taplamacioglu, M. C. (2023). Improvement of power transformer fault diagnosis by using sequential Kalman filter sensor fusion. International Journal of Electrical Power & Energy Systems, 149, 109038.
[11] Shi, X., Li, T., Fang, F., et al. (2024). Dissolved Gas Analysis for Power Transformer Fault Diagnosis Combining Domain Knowledge and Capsule Network. IEEE Transactions on Dielectrics and Electrical Insulation (Early Access). DOI: 10.1109/TDEI.2024.3374246.
[12] Li, Y., Li, Z., Chen, L. (2019). Dynamic state estimation of generators under cyber attacks. IEEE Access, 7, 125253-125267.
[13] Li, Y., Wei, X., Li, Y., Dong, Z., & Shahidehpour, M. (2022). Detection of false data injection attacks in smart grid: A secure federated deep learning approach. IEEE Transactions on Smart Grid, 13(6), 4862-4872.
[14] Zhou, X., Feng, J., & Li, Y. (2021). Non-intrusive load decomposition based on CNN–LSTM hybrid deep learning model. Energy Reports, 7, 5762-5771.
[15] Li, Y., & Yang, Z. (2017). Application of EOS-ELM with binary Jaya-based feature selection to real-time transient stability assessment using PMU data. IEEE Access, 5, 23092-23101.
[16] Li, Y., Zhang, M., & Chen, C. (2022). A deep-learning intelligent system incorporating data augmentation for short-term voltage stability assessment of power systems. Applied Energy, 308, 118347.
[17] Wang, L., Littler, T., & Liu, X. (2023). Hybrid AI model for power transformer assessment using imbalanced DGA datasets. IET Renewable Power Generation, to be published. DOI: 10.1049/rpg2.12733.
[18] Kim, S., Park, J., Kim, W., Jo, S. H., & Youn, B. D. (2022). Learning from even a weak teacher: Bridging rule-based Duval method and a deep neural network for power transformer fault diagnosis. International Journal of Electrical Power & Energy Systems, 136, 107619.
[19] Lv, G., Cheng, H., Zhai, H., & Dong, L. (2005). Fault diagnosis of power transformer based on multi-layer SVM classifier. Electric power systems research, 75(1), 9-15.
[20] Shi, Z. B., Li, Y., Song, Y. F., & Yu, T. (2009). Fault diagnosis of transformer based on quantum-behaved particle swarm optimization-based least squares support vector machines. In 2009 International Conference on Information Engineering and Computer Science (pp. 1-4). IEEE.
[21] Li, Z., He, Y., Xing, Z., & Duan, J. (2022). Transformer fault diagnosis based on improved deep coupled dense convolutional neural network. Electric Power Systems Research, 209, 107969.
[22] Yan, G., Li, X., & Zhu, L. (2006). Transformer fault diagnosis based on fuzzy neural network and dissolved gas analysis. Proceedings of the 2006 IEEE International Conference on Mechatronics and Automation, 2006, 2584-2588.
[23] Wang, Z., Wang, Y., Yang, H., & Zhang, Q. (2019). Transformer fault diagnosis using deep learning and dissolved gas analysis. IEEE Transactions on Power Delivery, 34(6), 2461-2470.
[24] Illias, H. A., & Zhao Liang, W. (2018). Identification of transformer fault based on dissolved gas analysis using hybrid support vector machine-modified evolutionary particle swarm optimisation. PLoS One, 13(1), e0191366.
[25] Wu, Y., Sun, X., Zhang, Y., Zhong, X., & Cheng, L. (2021). A power transformer fault diagnosis method-based hybrid improved seagull optimization algorithm and support vector machine. Ieee Access, 10, 17268-17286.
[26] Fan, Q., Yu, F., & Xuan, M. (2021). Transformer fault diagnosis method based on improved whale optimization algorithm to optimize support vector machine. Energy Reports, 7, 856-866.
[27] Shi, Z. B., Li, Y., & Yu, T. (2009). Short-term load forecasting based on LS-SVM optimized by bacterial colony chemotaxis algorithm. In 2009 International Conference on Information and Multimedia Technology (pp. 306-309). IEEE.
[28] Shi, Z. B., & Li, Y. (2009). Fault diagnosis of power transformer using LS-SVMs with BCC. In 2009 2nd IEEE International Conference on Computer Science and Information Technology (pp. 417-420). IEEE.

[29] Babnik, T., Aggarwal, R. K., & Moore, P. J. (2008). Principal component and hierarchical cluster analyses as applied to transformer partial discharge data with particular reference to transformer condition monitoring. IEEE transactions on power delivery, 23(4), 2008-2016.
[30] Miranda, V., & Castro, A. R. G. (2005). Improving the IEC table for transformer failure diagnosis with knowledge extraction from neural networks. IEEE transactions on power delivery, 20(4), 2509-2516.
[31] Li, Y., Li, G., & Wang, Z. (2015). Rule extraction based on extreme learning machine and an improved ant-miner algorithm for transient stability assessment. PloS one, 10(6), e0130814.
[32] Wang, L., Qu, Z., Li, Y., Hu, K., Sun, J., Xue, K., & Cui, M. (2020). Method for extracting patterns of coordinated network attacks on electric power CPS based on temporal–topological correlation. IEEE Access, 8, 57260-57272.
[33] Castro, A. R. G., & Miranda, V. (2005). An interpretation of neural networks as inference engines with application to transformer failure diagnosis. International Journal of Electrical Power & Energy Systems, 27(9-10), 620-626.
[34] Tightiz, L., Nasab, M. A., Yang, H., & Addeh, A. (2020). An intelligent system based on optimized ANFIS and association rules for power transformer fault diagnosis. ISA transactions, 103, 63-74.
[35] Kim, S., Park, J., Kim, W., Jo, S. H., & Youn, B. D. (2022). Learning from even a weak teacher: Bridging rule-based Duval method and a deep neural network for power transformer fault diagnosis. International Journal of Electrical Power & Energy Systems, 136, 107619.
[36] Lin, C. E., Ling, J. M., & Huang, C. L. (1993). An expert system for transformer fault diagnosis using dissolved gas analysis. IEEE transactions on Power Delivery, 8(1), 231-238.
[37]Peng, Z., & Song, B. (2009, May). Research on transformer fault diagnosis expert system based on DGA database. In 2009 Second International Conference on Information and Computing Science (Vol. 2, pp. 29-31). IEEE.
[38]Mani, G., & Jerome, J. (2014). Intuitionistic fuzzy expert system based fault diagnosis using dissolved gas analysis for power transformer. Journal of Electrical Engineering and Technology, 9(6), 2058-2064.
[39]Nagpal, T., & Brar, Y. S. (2015). Expert system based fault detection of power transformer. Journal of Computational and Theoretical Nanoscience, 12(2), 208-214.
[40]Xu, Y., Li, Y., Wang, Y., Zhong, D., & Zhang, G. (2021). Improved few-shot learning method for transformer fault diagnosis based on approximation space and belief functions. Expert Systems with Applications, 167, 114105.
[41] Senoussaoui, M. E. A., Brahami, M., & Fofana, I. (2018). Combining and comparing various machine‐learning algorithms to improve dissolved gas analysis interpretation. IET Generation, Transmission & Distribution, 12(15), 3673-3679.
[42] Liu, C., Cui, H., & Li, G. (2017). Fault Diagnosis for the Power Transformer Based on Multi-feature Fusion algorithm. In 2017 5th International Conference on Mechatronics, Materials, Chemistry and Computer Engineering (ICMMCCE 2017) (pp. 647-651). Atlantis Press.
[43] Kari, T., Gao, W., Zhao, D., Abiderexiti, K., Mo, W., Wang, Y., & Luan, L. (2018). Hybrid feature selection approach for power transformer fault diagnosis based on support vector machine and genetic algorithm. IET Generation, Transmission & Distribution, 12(21), 5672-5680.
[44] Sun, H. C., Huang, Y. C., & Huang, C. M. (2012). Fault diagnosis of power transformers using computational intelligence: A review. Energy Procedia, 14, 1226-1231.
[45] Rao, U. M., Fofana, I., Rajesh, K. N. V. P. S., & Picher, P. (2021). Identification and application of machine learning algorithms for transformer dissolved gas analysis. IEEE Transactions on Dielectrics and Electrical Insulation, 28(5), 1828-1835.
[46] Li, Z., He, Y., Xing, Z., & Duan, J. (2022). Transformer fault diagnosis based on improved deep coupled dense convolutional neural network. Electric Power Systems Research, 209, 107969.
[47] Shi, Z., Tong, Y., Chen, D., & Li, Y. (2009). Identification of beef freshness with electronic nose. Transactions of the Chinese Society for Agricultural Machinery, 40(11), 184-188.
[48] Gu, X., Li, Y., & Jia, J. (2015). Feature selection for transient stability assessment based on kernelized fuzzy rough sets and memetic algorithm. International Journal of Electrical Power & Energy Systems, 64, 664-670.
[49] Shi, Z. B., Yu, T., Zhao, Q., Li, Y., & Lan, Y. B. (2008). Comparison of algorithms for an electronic nose in identifying liquors. Journal of Bionic Engineering, 5(3), 253-257.
[50] Seifeddine, S., Khmais, B., & Abdelkader, C. (2012). Power transformer fault diagnosis based on dissolved gas analysis by artificial neural network. In 2012 First International Conference on Renewable Energies and Vehicular Technology (pp. 230-236). IEEE.

[51] Ghoneim, S. S., Taha, I. B., & Elkalashy, N. I. (2016). Integrated ANN-based proactive fault diagnostic scheme for power transformers using dissolved gas analysis. IEEE Transactions on Dielectrics and Electrical Insulation, 23(3), 1838-1845.
[52] Dai, J., Song, H., Sheng, G., & Jiang, X. (2017). Dissolved gas analysis of insulating oil for power transformer fault diagnosis with deep belief network. IEEE Transactions on Dielectrics and Electrical Insulation, 24(5), 2828-2835.
[53] Souahlia, S., Bacha, K., & Chaari, A. (2012). MLP neural network-based decision for power transformers fault diagnosis using an improved combination of Rogers and Doernenburg ratios DGA. International Journal of Electrical Power & Energy Systems, 43(1), 1346-1353.
[54] Sutikno, H., Prasojo, R. A., Abu-Siada, A. (2024). Machine learning based multi-method interpretation to enhance dissolved gas analysis for power transformer fault diagnosis. Heliyon, 10(4), e25975.
[55] Zhou, Y., Tao, L., Yang, X., & Yang, L. (2021). Novel probabilistic neural network models combined with dissolved gas analysis for fault diagnosis of oil-immersed power transformers. ACS Omega, 6(28), 18084-18098.
[56] Tao, L., Yang, X., Zhou, Y., & Yang, L. (2021). A novel transformers fault diagnosis method based on probabilistic neural network and bio-inspired optimizer. Sensors, 21(11), 3623.
[57] Abu-Siada, A., & Hmood, S. (2015). A new fuzzy logic approach to identify power transformer criticality using dissolved gas-in-oil analysis. International Journal of Electrical Power & Energy Systems, 67, 401-408.
[58] Huang, Y. C., Huang, C. M., & Sun, H. C. (2012). Data mining for oil‐insulated power transformers: an advanced literature survey. Wiley Interdisciplinary Reviews: Data Mining and Knowledge Discovery, 2(2), 138-148.
[59] Fang, Z., Zhao, D., Chen, C., Li, Y., & Tian, Y. (2020). Nonintrusive appliance identification with appliance-specific networks. IEEE Transactions on Industry Applications, 56(4), 3443-3452.
[60] Zhang, M., Li, J., Li, Y., & Xu, R. (2021). Deep learning for short-term voltage stability assessment of power systems. IEEE Access, 9, 29711-29718.
[61] Li, Y., He, S., Li, Y., Ge, L., Lou, S., & Zeng, Z. (2022). Probabilistic charging power forecast of EVCS: Reinforcement learning assisted deep learning approach. IEEE Transactions on Intelligent Vehicles.
[62] Islam, S. M., Wu, T., & Ledwich, G. (2000). A novel fuzzy logic approach to transformer fault diagnosis. IEEE Transactions on Dielectrics and electrical Insulation, 7(2), 177-186.
[63] Demirci, M., Gözde, H., & Taplamacioglu, M. C. (2023). Improvement of power transformer fault diagnosis by using sequential Kalman filter sensor fusion. International Journal of Electrical Power & Energy Systems, 149, 109038.
[64] Han, X., Huang, S., Zhang, X., et al. (2023). A transformer condition recognition method based on dissolved gas analysis features selection and multiple models fusion. Engineering Applications of Artificial Intelligence, 123, 106518.
[65] Li, Y., Li, G., Wang, Z., Han, Z., & Bai, X. (2015). A multifeature fusion approach for power system transient stability assessment using PMU data. Mathematical Problems in Engineering, 2015.
[66] Wani, S. A., Rana, A. S., Sohail, S., Rahman, O., Parveen, S., & Khan, S. A. (2021). Advances in DGA based condition monitoring of transformers: A review. Renewable and Sustainable Energy Reviews, 149, 111347.
[67] Hou, Y., Jia, S., Lun, X., Shi, Y., & Li, Y. (2020). Deep feature mining via attention-based BiLSTM-GCN for human motor imagery recognition. arXiv preprint arXiv:2005.00777.
[68] Mao, W., Wei, B., Xu, X., Chen, L., Wu, T., Peng, Z., & Ren, C. (2022). Fault Diagnosis for Power Transformers through Semi-Supervised Transfer Learning. Sensors, 22(12), 4470.
[69] Pei, X., Zheng, X., & Wu, J. (2021). Rotating machinery fault diagnosis through a transformer convolution network subjected to transfer learning. IEEE Transactions on Instrumentation and Measurement, 70, 1-11.
[70] Zhao, X., Chen, S., Gao, K., & Luo, L. (2023). Bidirectional Recurrent Neural Network based on Multi-Kernel Learning Support Vector Machine for Transformer Fault Diagnosis. International Journal of Advanced Computer Science and Applications, 14(1).
[71] Duval, M. (2008). The Duval triangle for load tap changers, non-mineral oils and low temperature faults in transformers. IEEE Electrical Insulation Magazine, 24(6), 22-29.

[72] Irungu, G. K., Akumu, A. O., & Munda, J. L. (2016). A new fault diagnostic technique in oil-filled electrical equipment; the dual of Duval triangle. IEEE Transactions on Dielectrics and Electrical Insulation, 23(6), 3405-3410.

[73] Wong, S. Y., Ye, X., Guo, F., & Goh, H. H. (2022). Computational intelligence for preventive maintenance of power transformers. Applied Soft Computing, 114, 108129.

[74] Mogos, A. S., Liang, X., & Chung, C. Y. (2023). Distribution transformer failure prediction for predictive maintenance using hybrid one-class deep SVDD classification and lightning strike failures data. IEEE Transactions on Power Delivery, to be published. DOI: 10.1109/TPWRD.2023.3268248

[75] Li, Y., Han, M., Yang, Z., et al. (2021). Coordinating flexible demand response and renewable uncertainties for scheduling of community integrated energy systems with an electric vehicle charging station: A bi-level approach. IEEE Transactions on Sustainable Energy, 12(4), 2321-2331.

[76] Kari, T., Gao, W., Zhao, D., Zhang, Z., Mo, W., Wang, Y., & Luan, L. (2018). An integrated method of ANFIS and Dempster-Shafer theory for fault diagnosis of power transformer. IEEE Transactions on Dielectrics and Electrical Insulation, 25(1), 360-371.

[77] Youssef, M. M., Ibrahim, R. A., Desouki, H., & Moustafa, M. M. Z. (2022, March). An overview on condition monitoring & health assessment techniques for distribution transformers. In 2022 6th International Conference on Green Energy and Applications (ICGEA) (pp. 187-192). IEEE.

[78] Li, Y., Li, K., Yang, Z., et al. (2022). Stochastic optimal scheduling of demand response-enabled microgrids with renewable generations: An analytical-heuristic approach. Journal of Cleaner Production, 330, 129840.

[79] Ali, M. S., Omar, A., Jaafar, A. S. A., et al. (2023). Conventional methods of dissolved gas analysis using oil-immersed power transformer for fault diagnosis: A review. Electric Power Systems Research, 216, 109064.

[80] Jin, Y., Wu, H., Zheng, J., et al. (2023). Power transformer fault diagnosis based on improved BP neural network. Electronics, 12(16), 3526.

[81] Li, Y., Wang, B., Yang, Z., et al. (2022). Optimal scheduling of integrated demand response-enabled community-integrated energy systems in uncertain environments. IEEE Transactions on Industry Applications, 58(2), 2640-2651.

[82] Li, Y., Han, M., Shahidehpour, M., et al. (2023). Data-driven distributionally robust scheduling of community integrated energy systems with uncertain renewable generations considering integrated demand response. Applied Energy, 335, 120749.

[83] Li, G., Zhai, X., Li, Y., Feng, B., Wang, Z., & Zhang, M. (2016). Multi-objective optimization operation considering environment benefits and economy based on ant colony optimization for isolated micro-grids. Energy Procedia, 104, 21-26.

[84] Chen, L., Jin, P., Yang, J., Li, Y., & Song, Y. (2021). Robust Kalman filter-based dynamic state estimation of natural gas pipeline networks. Mathematical Problems in Engineering, 2021, 1-10.

[85] Li, Y., Yang, Z., Li, G., Zhao, D., & Tian, W. (2018). Optimal scheduling of an isolated microgrid with battery storage considering load and renewable generation uncertainties. IEEE Transactions on Industrial Electronics, 66(2), 1565-1575.

[86] Christina, A., Salam, M. A., Rahman, Q. M., Wen, F., Ang, S. P., & Voon, W. (2018). Causes of transformer failures and diagnostic methods–A review. Renewable and Sustainable Energy Reviews, 82, 1442-1456.

[87] Li, Y., Wang, J., Zhao, D., Li, G., & Chen, C. (2018). A two-stage approach for combined heat and power economic emission dispatch: Combining multi-objective optimization with integrated decision making. Energy, 162, 237-254.

[88] de Faria Jr, H., Costa, J. G. S., & Olivas, J. L. M. (2015). A review of monitoring methods for predictive maintenance of electric power transformers based on dissolved gas analysis. Renewable and sustainable energy reviews, 46, 201-209.

[89] Ward, S. A., El-Faraskoury, A., Badawi, M., Ibrahim, S. A., Mahmoud, K., Lehtonen, M., & Darwish, M. M. (2021). Towards precise interpretation of oil transformers via novel combined techniques based on DGA and partial discharge sensors. Sensors, 21(6), 2223.

[90] Wani, S. A., Rana, A. S., Sohail, S., Rahman, O., Parveen, S., & Khan, S. A. (2021). Advances in DGA based condition monitoring of transformers: A review. Renewable and Sustainable Energy Reviews, 149, 111347.

[91] Yang, R., & Li, Y. (2022). Resilience assessment and improvement for electric power transmission systems against typhoon disasters: a data-model hybrid driven approach. Energy Reports, 8, 10923-10936.
[92] Xia, L., Liang, Y., Zheng, P., & Huang, X. (2022). Residual-hypergraph convolution network: A model-based and data-driven integrated approach for fault diagnosis in complex equipment. IEEE Transactions on Instrumentation and Measurement, 72, 3501811.
[93] Wang, G., Liu, Y., Chen, X., Yan, Q., Sui, H., Ma, C., & Zhang, J. (2021). Power transformer fault diagnosis system based on Internet of Things. Eurasip Journal on Wireless Communications and Networking, 2021, 1-24.
[94] Mohamad, A. A., Mezaal, Y. S., & Abdulkareem, S. F. (2018). Computerized power transformer monitoring based on internet of things. International Journal of Engineering & Technology, 7(4), 2773-2778.
[95] Elmashtoly, A. M., & Chang, C. K. (2020). Prognostics health management system for power transformer with IEC61850 and internet of things. Journal of Electrical Engineering & Technology, 15(2), 673-683.
[96] Jia, J., Tao, F., Zhang, G., Shao, J., Zhang, X., & Wang, B. (2020). Validity evaluation of transformer DGA online monitoring data in grid edge systems. IEEE Access, 8, 60759-60768.
[97] Ahmad, I., Singh, Y., & Ahamad, J. (2020). Machine learning based transformer health monitoring using IoT Edge computing. In 2020 5th International conference on computing, communication and security (ICCCS) (pp. 1-5). IEEE.
[98] Li, Y., Wang, R., & Yang, Z. (2021). Optimal scheduling of isolated microgrids using automated reinforcement learning-based multi-period forecasting. IEEE Transactions on Sustainable Energy, 13(1), 159-169.
[99] Zhang, Y., Li, T., Na, G., Li, G., & Li, Y. (2015). Optimized extreme learning machine for power system transient stability prediction using synchrophasors. Mathematical Problems in Engineering, 2015.
[100] Li, Y., Li, Y., Li, G., Zhao, D., & Chen, C. (2018). Two-stage multi-objective OPF for AC/DC grids with VSC-HVDC: Incorporating decisions analysis into optimization process. Energy, 147, 286-296.
[101] Li, Y., Feng, B., Wang, B., & Sun, S. (2022). Joint planning of distributed generations and energy storage in active distribution networks: A Bi-Level programming approach. Energy, 245, 123226.
[102] Liu, F., Li, Y., Li, B., Li, J., & Xie, H. (2021). Bitcoin transaction strategy construction based on deep reinforcement learning. Applied Soft Computing, 113, 107952.
[103] Li, Y., Bu, F., Li, Y., & Long, C. (2023). Optimal scheduling of island integrated energy systems considering multi-uncertainties and hydrothermal simultaneous transmission: A deep reinforcement learning approach. Applied Energy, 333, 120540.
[104] Hong, K., Jin, M., & Huang, H. (2020). Transformer winding fault diagnosis using vibration image and deep learning. IEEE Transactions on Power Delivery, 36(2), 676-685.
[105] Lin, J., Ma, J., & Zhu, J. (2022). Hierarchical federated learning for power transformer fault diagnosis. IEEE Transactions on Instrumentation and Measurement, 71, 1-11.
[106] Li, Y., Li, J., & Wang, Y. (2021). Privacy-preserving spatiotemporal scenario generation of renewable energies: A federated deep generative learning approach. IEEE Transactions on Industrial Informatics, 18(4), 2310-2320.
[107] Li, Y., Wang, R., Li, Y., Zhang, M., & Long, C. (2023). Wind power forecasting considering data privacy protection: A federated deep reinforcement learning approach. Applied Energy, 329, 120291.
[108] Singh, P., Masud, M., Hossain, M. S., Kaur, A., Muhammad, G., & Ghoneim, A. (2021). Privacy-preserving serverless computing using federated learning for smart grids. IEEE Transactions on Industrial Informatics, 18(11), 7843-7852.
[109] Li, Y., He, S., Li, Y., Shi, Y., & Zeng, Z. (2023). Federated multiagent deep reinforcement learning approach via physics-informed reward for multimicrogrid energy management. IEEE Transactions on Neural Networks and Learning Systems, to be published. DOI: 10.1109/TNNLS.2022.3232630.
[100] Li, L., Fan, Y., Tse, M., & Lin, K. Y. (2020). A review of applications in federated learning. Computers & Industrial Engineering, 149, 106854.
[111] Manoj, T., Ranga, C., Ghoneim, S. S., et al. (2023). Alternate and effective dissolved gas interpretation to understand the transformer incipient faults. IEEE Transactions on Dielectrics and Electrical Insulation, 30(3), 1231-1239.

[112] Rudin, C. (2019). Stop explaining black box machine learning models for high stakes decisions and use interpretable models instead. Nature machine intelligence, 1(5), 206-215.
[113] Saroja, S., Haseena, S., Madavan, R. (2023). Dissolved gas analysis of transformer: An approach based on ML and MCDM. IEEE Transactions on Dielectrics and Electrical Insulation, 30(5), 2429-2438.
[114] Guerbas, F., Benmahamed, Y., Teguar, Y., et al. (2024). Neural networks and particle swarm for transformer oil diagnosis by dissolved gas analysis. Scientific Reports, 14(1), 9271.